\documentclass[
fleqn,usenatbib]{mnras}

\usepackage[T1]{fontenc}
\usepackage{ae,aecompl}

\usepackage{graphicx}	
\usepackage{amsmath}	
\usepackage{amssymb}	
\usepackage{multirow}   

\allowdisplaybreaks[4] 

\title[Effects of $H_0$ on $f\sigma_8$ tension]{Testing the effect of $H_0$ on $f\sigma_8$ tension using a Gaussian Process method}

\author[E. K. Li et al.]{%
En-Kun Li,\thanks{E-mail: ekli\_091@mail.dlut.edu.cn}
Minghui Du,\thanks{E-mail: angelbeats@mail.dlut.edu.cn}
Zhi-Huan Zhou,\thanks{E-mail: scott@mail.dlut.edu.cn}
Hongchao Zhang,\thanks{E-mail: zhanghc@mail.dlut.edu.cn}
Lixin Xu\thanks{Corresponding author: lxxu@dlut.edu.cn}
\\
Institute of Theoretical Physics, School of Physics, Dalian University of Technology, Dalian, 116024, China
}

\date{Accepted XXX. Received YYY; in original form ZZZ}

\pubyear{2020}

\begin{document}
\label{firstpage}
\pagerange{\pageref{firstpage}--\pageref{lastpage}}
\maketitle

\begin{abstract}
Using the $f\sigma_8(z)$ redshift space distortion (RSD) data, the $\sigma_8^0-\Omega_m^0$ tension is studied utilizing a parameterization of growth rate $f(z) = \Omega_m(z)^\gamma$.
Here, $f(z)$ is derived from the expansion history $H(z)$ which is reconstructed from the observational Hubble data applying the Gaussian Process method.
It is found that different priors of $H_0$ have great influences on the evolution curve of $H(z)$ and the constraint of $\sigma_8^0-\Omega_m^0$.
When using a larger $H_0$ prior, the low redshifts $H(z)$ deviate significantly from that of the $\Lambda$CDM model, which indicates that a dark energy model different from the cosmological constant can help to relax the $H_0$ tension problem.
The tension between our best-fit values of $\sigma_8^0-\Omega_m^0$ and that of the \textit{Planck} 2018 $\Lambda$CDM (PLA) will disappear (less than $1\sigma$) when taking a prior for $H_0$ obtained from PLA.
Moreover, the tension exceeds $2\sigma$ level when applying the prior  $H_0 = 73.52 \pm 1.62$ km/s/Mpc resulted from the Hubble Space Telescope photometry.
By comparing the $S_8 -\Omega_m^0$ planes of our method with the results from KV450+DES-Y1, we find that using our method and applying the RSD data may be helpful to break the parameter degeneracies.
\end{abstract}




\section{Introduction}
\label{sec:Intro}

The past and present analyses of various cosmological observations converge to the fact that our Universe is undergoing an accelerated expansion phase \citep{Hicken:2009dk, Komatsu:2010fb, Blake:2011en,Hinshaw:2012aka, Farooq:2012ev, Ade:2015xua, Aghanim:2018eyx}.
To explain this phenomenon, two kinds of interpretations have been raised.
One is proposing an unknown component with negative pressure called dark energy in the context of General Relativity (GR), and the other is modifying the laws of gravity (MG).
Based on these two branches, numerous models have been presented.
Among these models, Lambda Cold Dark Matter ($\Lambda$CDM) model is the most simple one and can excellently fit with almost all observational data, such as the cosmic microwave background (CMB) radiations \citep{Hinshaw:2012aka, Ade:2015xua, Aghanim:2018eyx}, the baryon acoustic oscillations (BAO) \citep{Percival:2007yw, Delubac:2014aqe, Alam:2016hwk, Percival:2018twa}, and the type Ia Supernovae (SNIa) \citep{Perlmutter:1998np, Riess:1998cb, Suzuki:2011hu, Betoule:2014frx}, etc.

Nonetheless, it is becoming exceedingly apparent that there are some discrepancies between the \textit{Planck} $\Lambda$CDM results and some independent observations in intermediate cosmological scale \citep{Raveri:2015maa}.
These discrepancies include the estimates of the Hubble constant $H_0$ \citep{Abdalla:2014cla, Bernal:2016gxb, Qing-Guo:2016ykt, Sola:2016jky, Riess:2018byc, Riess:2019cxk}, the matter density parameter $\Omega_m^0$ and the amplitude of the power spectrum on the scale of $8h^{-1}$Mpc ($\sigma_8^0$) \citep{Gao:2013pfa, Battye:2014qga, Bull:2015stt, Sola:2017jbl, McCarthy:2017csu}, etc.
In order to solve these discrepancies, different methods and cosmology models have been reported, including viscous bulk cosmology \citep{Mostaghel:2016lcd}, assuming a variable Newton constant \citep{Nesseris:2017vor, Kazantzidis:2018rnb}, considering the interaction between neutrinos and dark matter \citep{DiValentino:2017oaw}, introducing interacting dark energy \citep{DiValentino:2017iww, Yang:2018euj}, model-independent method \citep{Zhao:2017cud, Gomez-Valent:2018hwc, Li:2019qic}, and so on.

Precise large-scale structure measurements are helpful to distinguish different models because these models may have different growth histories of structure.
As a starting point, in the subhorizon ($k \gg aH$), the equation that describes the evolution of the linear matter growth factor $\delta = \delta \rho_m/\rho_m$ in the context of GR and most MG models has the form \citep{Kazantzidis:2018rnb}
\begin{equation}
\delta^{\prime\prime} + \left(\frac{H^\prime}{H} -\frac{1}{1+z} \right) \delta^\prime \simeq \frac{3}{2} (1+z) \frac{H_0^2}{H^2} \frac{G_{\text{eff}}(z,k)}{G_N} \Omega_m^0 \delta,
\label{eq:dd_delta}
\end{equation}
where $\rho_m$ is the background matter density, $H = \dot{a}/a$ is the Hubble expansion rate at scale factor $a=1/(1+z)$, $G_N$ is the Newton's constant, $G_{\text{eff}}$ is the effective Newton's constant which in general may depend on both the redshift $z$ and the cosmological scale $k$, and `` $\prime$ '' denotes a derivative with respect to $z$.
In GR we have $G_{\text{eff}} =G_N$ while in MG $G_{\text{eff}}/G_N$ may vary with both cosmological redshift and scale.

Although it is difficult to give the analytical solution of Eq.~\eqref{eq:dd_delta}, a good parameterization of the growth rate $f(a) \equiv d\ln \delta/d\ln a$ is given by \citep{Wang:1998gt, Amendola:2004wa, Linder:2005in, Polarski:2007rr, Gannouji:2008jr, Polarski:2016ieb, Nesseris:2015fqa}
\begin{equation}
f(a) \simeq \Omega_m(a)^\gamma,
\label{eq:fz}
\end{equation}
where $\Omega_m(a) = \Omega_m^0 a^{-3} /(H(a)/H_0)^{2}$ is the fractional matter density, and $\gamma$ is the growth index.
The growth index differs between different cosmological models \citep{Xu:2013tsa, Polarski:2016ieb}.
In the $\Lambda$CDM model, $\gamma = 6/11$ is a solution to Eq.~\eqref{eq:dd_delta} where the terms $\mathcal{O}(1-\Omega_m(a))^2$ are neglected \citep{Wang:1998gt}, while $\gamma \simeq 0.55$ is that of dark energy models with slowly varying equation of state \citep{Linder:2005in}.
For MG models, different values are predicted, e.g., $\gamma \simeq 0.68$ for Dvali-Gabadadze-Porrati (DGP) braneworld model \citep{Linder:2007hg, Wei:2008ig}.
Applying some model-independent methods, authors in Refs.~\citep{LHuillier:2017ani, Shafieloo:2018gin, LHuillier:2019imn} found that the value of $\gamma$ is consistent with that of the flat-$\Lambda$CDM model, and Yin \& Wei \citep{Yin:2018mvu, Yin:2019rgm} also investigated the time varying $\gamma(z)$.
Since most of the information on linear clustering is expected to come from the epoch of equality of matter and dark energy, it is reasonable to use this parameterization to approximate $f(z)$ \citep{Pogosian:2010tj}.

In particular, most of the growth rate measurements can be obtained from redshift space distortion (RSD) measurements via the peculiar velocities of galaxies \citep{Kaiser:1987qv}.
However, $f(z)$ is sensitive to the bias parameter $b$, which makes the observation of $f(z)$ data unreliable \citep{Nesseris:2017vor}.
Therefore, most growth rate measurements are reported as the combination $f(z)\sigma_8(z) = f\sigma_8(z)$ instead of $f(z)$, where $\sigma_8(z) = \sigma_8^0 \delta(z)/\delta_0$ is the matter power spectrum normalization on scales of $8h^{-1}$Mpc.
In addition, the joint measurement of expansion history and growth history provides an important test of GR and can help to break the degeneracies between MG theories and dark energy models in GR \citep{Linder:2005in, Linder:2016xer}.
In this paper, using the RSD data and the observational Hubble data (OHD), we will investigate the $\sigma_8^0-\Omega_m^0$ tension utilizing the Gaussian Process method.
We reconstruct the expansion history $H(z)$ firstly using the OHD data with priors for Hubble constant $H_0$, and then derive the theoretical value of $f\sigma_8(z)$ applying the parameterization $f(z) = \Omega_m(z)^\gamma$. 
Finally, by adopting the Markov Chain Monte Carlo (MCMC) method, the constraints on the free parameters in $f\sigma_8(z)$ are given using the RSD data corrected by the fiducial model corrections.

The layout of this work is as follows.
In section~\ref{sec:meth_data}, we introduce the basic methodology adopted to derive $f\sigma_8(z)$ and the observational data combinations used to constrain free parameters.
And then, in section~\ref{sec:reconst}, we show the reconstruction of Hubble parameter $H(z)$ and $f\sigma_8(z)$ under different combinations of OHD.
Our results and discussions are displayed in section~\ref{sec:result}.
At last, we summarize our conclusions in section~\ref{sec:sum}.

\section{Methodology and observational data}
\label{sec:meth_data}

\subsection{Methodology}
\label{sec:fsigma_theory}

As we known that most growth rate measurements are reported as the combination $f\sigma_8(z)$.
Using the definitions of $f(z)$, $\sigma_8(z)$ and Eq.~\eqref{eq:fz}, one can obtain
\begin{align}
f\sigma_8(z) &= \sigma_8^0 \left(\Omega_m^0 H_0^2 \right)^\gamma \left( \frac{(1+z)^3}{H(z)^2} \right)^\gamma \nonumber \\
 & \cdot \exp\left[ -\left( \Omega_m^0 H_0^2 \right)^\gamma \int_0^z \frac{(1+z')^{3\gamma-1} }{H(z')^{2\gamma}} dz' \right].
\label{eq:fsigma_8}
\end{align}
Thus, given an expansion history function $H(z)$ or $H(z)/H_0$, we can reconstruct the observable quantity $f\sigma_8(z)$, assuming $\sigma_8^0$, $\Omega_m^0$ and $\gamma$ are known.

The Gaussian Process method \citep{rasmussen2006gaussian, Seikel:2012uu, Seikel:2013fda} can provide a smooth reconstructed $H(z)$ using the combination of OHD without assuming a parametrisation of the function.
So we can get a full model-independent reconstructed $f\sigma_8(z)$ with three free parameters $\{ \sigma_8^0, \Omega_m^0, \gamma \}$ using Eq.~\eqref{eq:fsigma_8}.

Now, we can use a $\chi^2$ minimization to constrain the three free parameters,
\begin{align}
\chi^2 &= \Delta  {\bf V^T} \textbf{Cov}^{-1} \Delta {\bf V}, \\
\Delta V_i &= f\sigma_{8,\text{obs}}(z_i) - f\sigma_{8,\text{rec}}(z_i) \\
\textbf{Cov} &= \textbf{Cov}^{\text{obs}} +\textbf{Cov}^{\text{rec}},
\label{eq:chisq}
\end{align}
where $\textbf{Cov}^{\text{obs}}$ is the covariance matrix of $f\sigma_{8,\text{obs}}$ and $\textbf{Cov}^{\text{rec}}$ is the covariance matrix of the reconstructed $f\sigma_{8,\text{rec}}(z)$ which is defined in Eq.~\eqref{eq:fsigma_8}.
The likelihood of the free parameters can be obtained from $\mathcal{L} \propto \exp[-\chi^2/2]$.
The constraints on the free parameters are performed using the Markov Chain Monte Carlo (MCMC) sampling method.
It's easy to do this by using the publicly available code \textbf{Cobaya}~\footnote{\url{https://github.com/CobayaSampler/cobaya}}, which calls the MCMC sampler developed for CosmoMC \citep{Lewis:2002ah,Lewis:2013hha}.

Furthermore, in order to quantify the tension between different estimate of parameter $\xi$, we need to introduce a quantization function of the tension level.
Assuming the $68\%$ confidence level ranges of parameter $\xi$ is $\xi \in [\xi_1 -\sigma_{1, \text{low}}, \xi_1 +\sigma_{1,\text{up}}]$ from observation $O_1$, and $\xi \in [\xi_2 -\sigma_{2, \text{low}}, \xi_2 +\sigma_{2,\text{up}}]$ from observation $O_2$.
Then, the simplest and most intuitive way to measure the degree of tension can be written as \citep{Zhao:2017jma}
\begin{equation}
s \equiv \frac{\xi_1 - \xi_2}{\sqrt{\sigma_{1, \text{low}}^2 + \sigma_{2, \text{up}}^2}},
\label{eq:tension}
\end{equation}
for the case $\xi_1 > \xi_2$.
This means that the tension of $\xi$ between $O_1$ and $O_2$ is at $s\sigma$ level.

\subsection{$f\sigma_8(z)$ measurements}
\label{sec:fsigma8_data}

Table~\ref{tab:fsigma8_data} shows a sample consisting of 63 observational $f\sigma_{8,\text{obs}}(z)$ RSD data points collected by 
\cite{Kazantzidis:2018rnb}.
It comprises the data published by various surveys from 2006 to the present and the parameters of the corresponding fiducial cosmology model are also shown in this table.
For more details please refer to Ref.~\citep{Kazantzidis:2018rnb} and references therein.

The covariance matrix of the 63 $f\sigma_8$ data points are assumed to be diagonal except for the WiggleZ subset of the data (three data points).
The covariance matrix of the three points of WiggleZ has been published as
\begin{equation}
C_{i,j}^{\text{WiggleZ}} = 10^{-3} \times
\begin{pmatrix}
6.400 & 2.570 & 0.000 \\
2.570 & 3.969 & 2.540 \\
0.000 & 2.540 & 5.184
\end{pmatrix}.
\label{eq:Cij_WiggleZ}
\end{equation}

One should note that all the $f\sigma_{8,\text{obs}}(z)$ data listed in Table~\ref{tab:fsigma8_data} are obtained assuming a fiducial $\Lambda$CDM cosmology \citep{Kazantzidis:2018rnb}.
Thus, the Alcock-Paczynski (AP) effect \citep{Alcock:1979mp} should be considered.
In the present paper, we will use the following rough approxmation of the AP effect \citep{Macaulay:2013swa, Kazantzidis:2018rnb}
\begin{equation}
f\sigma_{8,\text{ap}}(z) \simeq \frac{H(z) D_A(z)}{H^{\text{fid}}(z,\Omega_m) D_A^{\text{fid}}(z,\Omega_m) } f\sigma_{8,\text{obs}}(z),
\label{eq:AP}
\end{equation}
where $D_A(z)$ is the angular diameter distance, and it can be written as
\begin{equation}
D_A(z) = \frac{c}{1+z} \int_0^z \frac{dz'}{H(z')},
\label{eq:D_A}
\end{equation}
in the spatially flat universe.

\subsection{Observational Hubble data}
\label{sec:OHD_dat}

The Hubble parameter $H(z)$ is usually evaluated as a function of the redshift $z$
\begin{equation}
H(z)= - \frac{1}{1+z} \frac{dz}{dt}.
\label{eq:Hubble}
\end{equation}
It can be seen that $H(z)$ depends on the derivative of redshift with respect to cosmic time.
The $H(z)$ measurements can be obtained via two approaches.
One is calculating the differential ages of passively evolving galaxies \citep{Jimenez:2001gg} providing $H(z)$ measurements that are model-independent.
This method is usually called the cosmic chronometers (CC).
The other method is based on the clustering of galaxies or quasars, which is firstly proposed by \citep{Gaztanaga:2008xz}, where the BAO peak position is used as a standard ruler in the radial direction.

Here, we use the compilation of OHD data points collected by Magana, et al. \citep{Magana:2017nfs} and Geng, et al. \citep{Geng:2018pxk}, including almost all $H(z)$ data reported in various surveys so far.
The 31 CC $H(z)$ data points are listed in Table~\ref{tab:Hubble_CC} and the 23 $H(z)$ data points obtained from clustering measurements are listed in Table~\ref{tab:Hubble_BAO}.
One may find that some of the $H(z)$ data points from clustering measurements are correlated since they either belong to the same analysis or there is overlap between the galaxy samples.
Here in this paper, we mainly take the central value and standard deviation of the OHD data into consideration. Thus, just as in Ref.~\citep{Geng:2018pxk}, we assume that they are independent measurements.

In addition, there is no observation for $H_0$ in these OHD data points mentioned above, so we also consider two different priors of $H_0$.
One is $H_0 = 67.27 \pm 0.60 \text{~km/s/Mpc}$ \citep{Aghanim:2018eyx} provided by \textit{Planck 2018} power spectra (TT,TE,EE+lowE) measurements by assuming base $\Lambda$CDM model (hereafter $P18$).
The other is $H_0 = 73.52 \pm 1.62 \text{~km/s/Mpc}$ presented by the Hubble Space Telescope photometry of long-period Milky Way Cepheids and GAIA parallaxes \citep{Riess:2018byc} (hereafter $R18$).

\section{Model-independent reconstruction}
\label{sec:reconst}

GP method provides a technique to reconstruct a function using the observational data without assuming a specific parameterization.
It is easy to reconstruct the Hubble parameters directly from the OHD data applying a freely available GaPP package \footnote{\url{http://www.acgc.uct.ac.za/~seikel/GAPP/index.html}}.
The numerical program written by ourselves is used in this paper, and there is no difference between our code and that of GaPP, which also indicates our program is credible.

GP are characterized by mean and covariance functions, which are defined by a small number of hyperparameters.
Throughout this work, we assume a priori mean function equal to zero, and use the squared exponential covariance function:
\begin{equation}
k(x,y) = \sigma_f^2 \exp\left( -\frac{(x-y)^2}{2l^2} \right),
\label{eq:kernel}
\end{equation}
where $\sigma_f$ and $l$ are two hyperparameters which can be determined by the observational data.
Supposing an observational data-set $\{x_i, y_i, \sigma_i\}$, where $x_i$ is the location of data point $i$, $y_i=f(x_i) + \epsilon_i$ is the corresponding actual observed value which is assumed to be scattered around the underlying function $f(x_i)$ and Gaussian noise with variance $\sigma_i$ is assumed.
Using the GP method, the reconstructed mean value and covariance of the underlying function $f(x)$ can be written as \citep{Seikel:2012uu}
\begin{align}
f(x) &= \sum_{i,j=1}^N k(x,x_i) (M^{-1})_{ij} y_i, 
\label{eq:mean_f}\\
\text{Cov}(f_x,f_y) & = k(x,y) - \sum_{i,j=1}^N k(x,x_i) (M^{-1})_{ij} k(x_j,y),
\label{eq:cov_ff}
\end{align}
where $M_{ij} = k(x_i, x_j) +C_{ij}$ and $C_{ij}$ is the covariance matrix of the observational data.

\subsection{Reconstruction of Hubble parameter}
\label{sec:rec_hz}

\begin{figure*}
\centering
\includegraphics[width=0.73\linewidth]{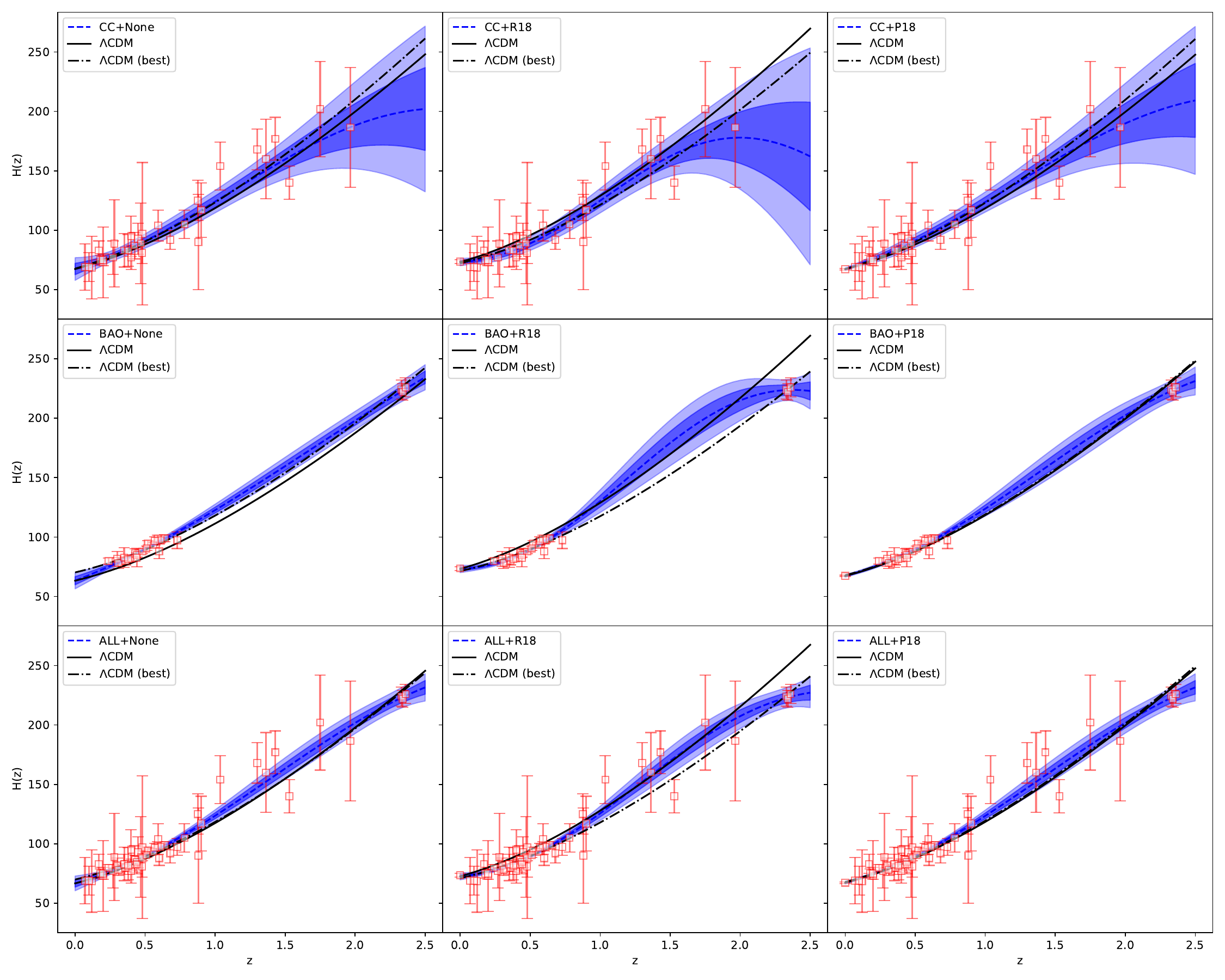}
\caption{The blue region is the reconstructed $H(z)$ as function of redshift $z$ within $2\sigma$ confidence level. The solid black lines represent the $\Lambda$CDM model with $\Omega_m = 0.3$ and $H_0$ is the reconstructed mean value from the corresponding data combinations. The dot dashed lines are the $\Lambda$CDM model using the mean values of the model parameters from Table~\ref{tab:results}. The red error bar plot is OHD data.}
\label{fig:rec_hz}
\end{figure*}

\begin{figure*}
\centering
\includegraphics[width=0.73\linewidth]{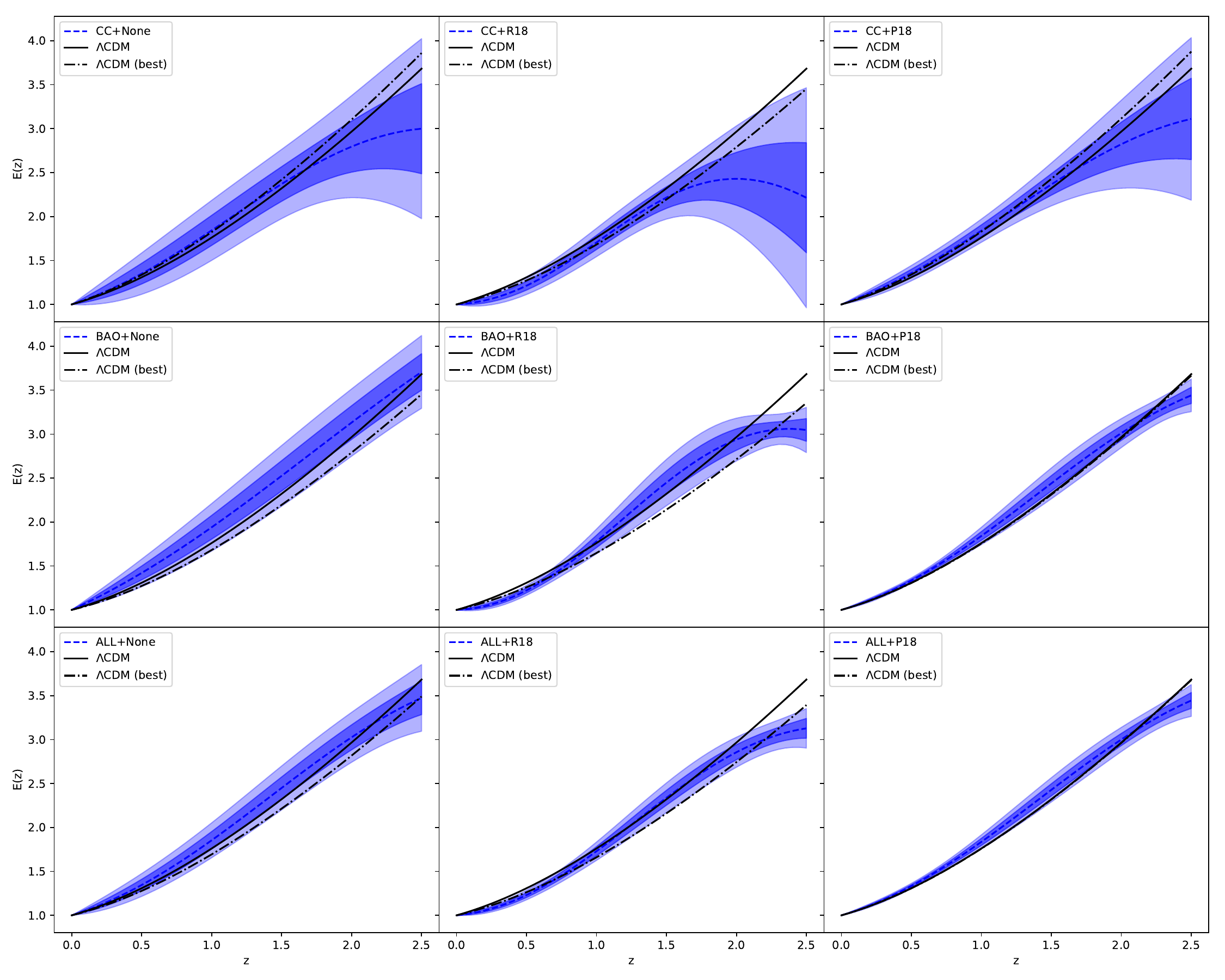}
\caption{The reconstructed $E(z)$ as function of redshift $z$ within $2\sigma$ region. The solid black lines are the $\Lambda$CDM model with $\Omega_m = 0.3$. The dot dashed lines are the $\Lambda$CDM model using the mean values of the model parameters from Table~\ref{tab:results}. }
\label{fig:rec_Ez}
\end{figure*}

\begin{figure*}
\centering
\includegraphics[width=0.73\linewidth]{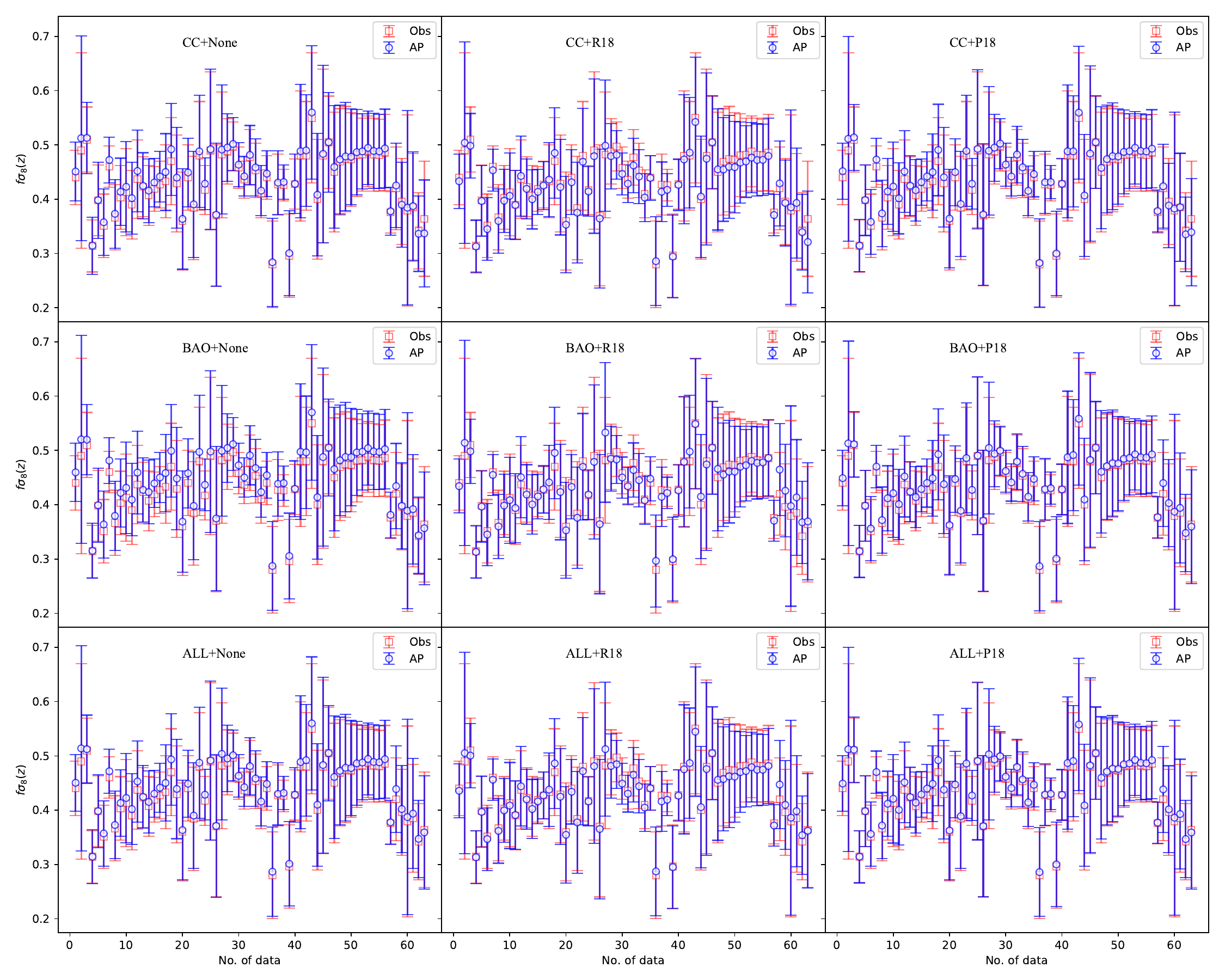}
\caption{Here the index ``Obs'' is for the original data and ``AP'' is the data after fiducial model correction. The $f\sigma_8^{ap}(z)$ data in different graphs are obtained with different reconstructions of $H(z)$. }
\label{fig:AP_fs8}
\end{figure*}

\begin{figure*}
\centering
\includegraphics[width=0.73\linewidth]{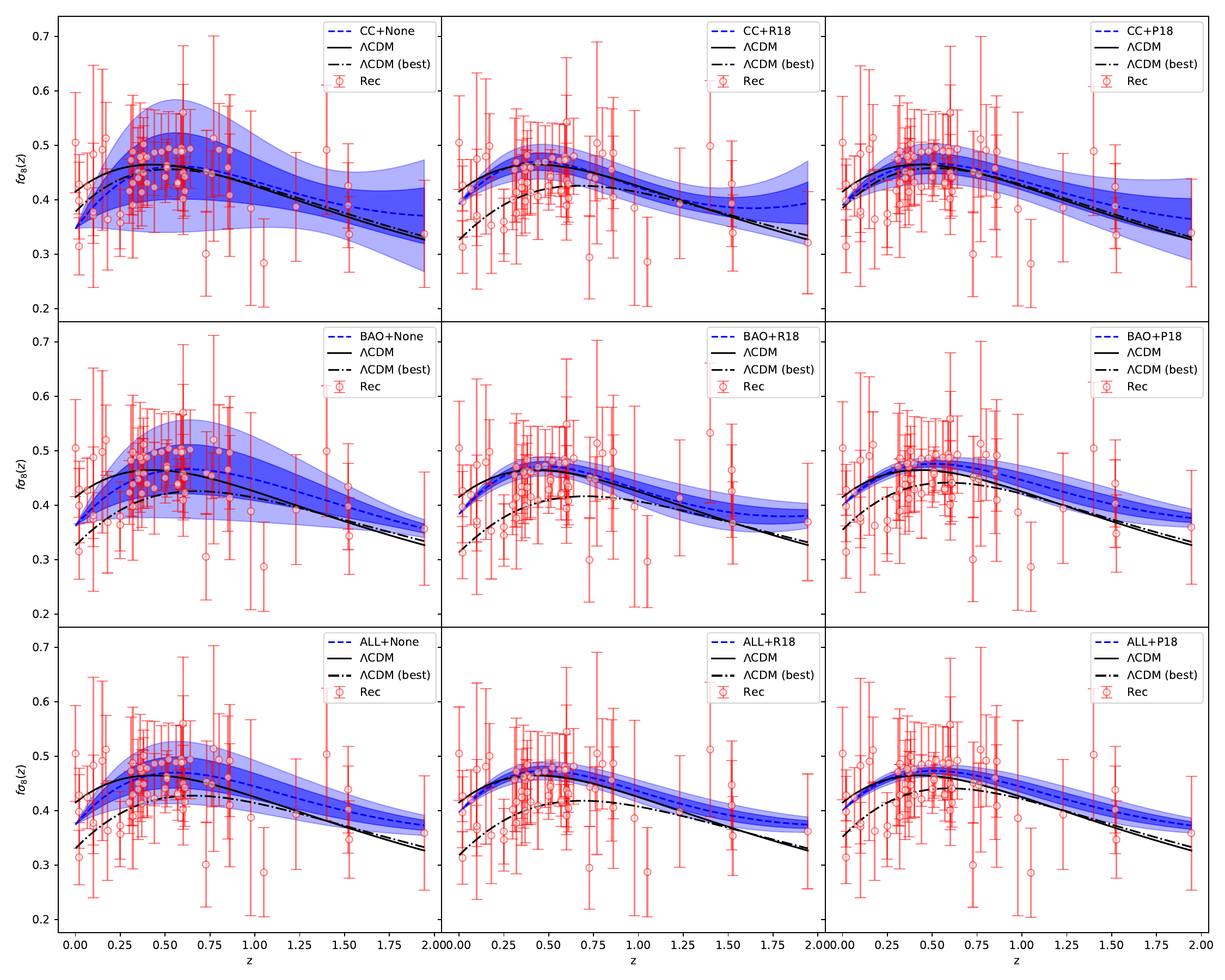}
\caption{The blue region is the reconstruction of $f\sigma_8(z)$ using the reconstructed $H(z)$ from OHD data through Eq.~\eqref{eq:rec_fs8_Ez} within $2\sigma$ region, the 
dot dashed lines are $f\sigma_8(z)$ derived form Eq.~\eqref{eq:fsigma_8} under $\Lambda$CDM model with different parameters. The red error bar plot are the measurements of RSD data after fiducial model correction. The parameters used here are the mean values from Table~\ref{tab:results}. For comparison, the $\Lambda$CDM model with the fixed parameters: $\Omega_m = 0.3$, $\sigma_8 = 0.8$, and $\gamma = 0.545$ are also shown in every panels (solid black lines).}
\label{fig:rec_fsig8}
\end{figure*}

By using Eqs.~\eqref{eq:mean_f} and \eqref{eq:cov_ff}, one can easily get the reconstructed Hubble parameters $H(z)$ and its covariance matrix between different redshifts.
The propagated covariance \citep{Alam:2004ip, Nesseris:2005ur, Wang:2010kwa, Xu:2013tsa} of the reconstructed $E(z)$ can be calculated with the reconstructed $H(z)$, and its covariance matrix is
\begin{align}
\text{Cov}[E_i, E_j] =& E_i E_i \left( \frac{\text{Cov}[H_i, H_j]}{H_i H_j} -\frac{\text{Cov}[H_i, H_0]}{H_i H_0}  \right. \nonumber \\
 & \left. -\frac{\text{Cov}[H_0, H_j]}{H_0 H_j}  + \frac{\text{Cov}[H_0, H_0]}{H_0 H_0} \right)
\end{align}
where $E_i = E(z_i) = H(z_i)/H_0$ is the dimensionless Hubble parameter, and $H_i = H(z_i)$ is the reconstructed Hubble parameter at $z_i$.

Next, we will examine the differences between the reconstructed $H(z)$ under various OHD data combinations.
In Figs.~\ref{fig:rec_hz} and \ref{fig:rec_Ez}, we have ploted the reconstructed $H(z)$ and $E(z)$ within $2\sigma$ region using different data combinations, and the reconstructed $H_0$ is also listed in Table~\ref{tab:results}.
Here the indexes CC, BAO and ALL represent the OHD data obtained from CC, clustering measurements, and CC+clustering measurements, respectively.
We also consider three different priors of $H_0$, i.e., no prior (index None), $H_0$ of R18 and $H_0$ of P18.
The three panels of the first column in Figs.~\ref{fig:rec_hz} and \ref{fig:rec_Ez} are the reconstructed results from the three data combinations with no prior on $H_0$.
From the values of $H_0$ listed in Table~\ref{tab:results}, one may find that OHD from BAO prefers a much smaller $H_0$ than P18 or R18, and the mean value of the derived $H_0$ from the CC data is much close to that of P18.
However, we should note that the tension level of the three reconstructed $H_0$ with P18 or R18 are all less than $2\sigma$ as a result of the big error bars. 

Comparing the three panels in the same rows of Figs.~\ref{fig:rec_hz} and \ref{fig:rec_Ez}, one can find that adding a prior on $H_0$ can significantly reduce the error bars of the reconstructed $H(z)$ at low redshifts, because the $H_0$ measurements of R18 and P18 have much smaller variances than the rest OHD data points.
It also can be found that the slope of the reconstructed $H(z)$ varies when choosing different priors of $H_0$, especially at low redshifts $z<0.5$.
For the cases of using the same OHD data points but with different $H_0$ priors, we find that the evolution curves of $H(z)$ under the P18 prior are similar to that without $H_0$ prior, this is due to the fact that they have similar mean values of $H_0$.

Meanwhile, from Figs.~\ref{fig:rec_hz} and \ref{fig:rec_Ez}, we can see that CC data gives a much looser reconstruction of $H(z)$ at higher redshifts than BAO data, which is because the BAO data points have much smaller variances at high redshifts.
And the $\Lambda$CDM model with fixed $\Omega_m$ and the mean values from the constraint results in Table~\ref{tab:results} are also shown in the two figures as comparison.
One can also find that when the R18 prior are took into consideration, our reconstructions are much different with the $\Lambda$CDM model.

\subsection{$f\sigma_8(z)$ data after fiducial model correction}
\label{sec:rec_fs8_ap}

In this work, we will use the reconstructed model-independent $H(z)$ and $D_A(z)$ for fiducial model correction. 
Thus, the central value of $f\sigma_{8,\text{ap}}(z)$ and its covariance matrix can be calculated according to Eq.~\eqref{eq:AP}.
The covariance matrix will be
\begin{equation}
\text{Cov}_{ij}^{ap} = q_i q_j \text{Cov}_{ij}^* + p_i p_j \text{Cov}_{ij}^{HD},
\label{eq:cov_ap}
\end{equation}
where $\text{Cov}^*$ and $\text{Cov}^{HD}$ are the covariance of observational RSD data and the reconstructed $H(z)D_A(z)$, respectively, $q_i = H_i D_{A,i}/(H_i^{\text{fid}} D_{A,i}^{fid})$, and $p_i = f\sigma_{8,i,\text{obs}}/(H_i^{\text{fid}} D_{A,i}^{fid})$.

The original $f\sigma_{8,\text{obs}}(z)$ data and the fiducial model correction data $f\sigma_{8,\text{ap}}(z)$ are ploted in Fig.~\ref{fig:AP_fs8}.
As shown in Fig.~\ref{fig:AP_fs8} this correction has little effect on the mean values of $f\sigma_8(z)$.
After some calculations, we find that the largest corrections on the mean values and the variances are less than $11\%$ and $18\%$, respectively.
The correlations between different data points also need to be taken into account when constraining on the free parameters.

\subsection{Reconstruction of $f\sigma_8(z)$}
\label{sec:rec_fs8_hz}

Using the reconstructed $H(z)$ or $E(z)$, the reconstructed $f\sigma_8(z)$ can be obtained through Eq.~\eqref{eq:fsigma_8}.
The mean value and the propagated covariance \citep{Alam:2004ip, Nesseris:2005ur, Wang:2010kwa, Xu:2013tsa} of the reconstructed $f\sigma_8(z)$ can be written as
\begin{align}
& f\sigma_8(z_i) = \sigma_8^0 (\Omega_m^0)^\gamma \frac{(1+z)^{3\gamma}}{E^{2\gamma}(z)} \exp\left[- (\Omega_m^0)^\gamma I_i \right],  \label{eq:rec_fs8_Ez} \\
& \text{Cov}[f\sigma_{8,i}, f\sigma_{8,j}] = 4\gamma^2 f\sigma_{8,i}, f\sigma_{8,j} \bigg\{ \frac{\text{Cov}[E_i, E_j]}{ E_i E_j } \nonumber \\
&\qquad\qquad - (\Omega_m^0)^\gamma M_{ij} - (\Omega_m^0)^\gamma M_{ji} + (\Omega_m^0)^{2 \gamma} N_{ij}  \bigg\},
\label{eq:rec_fs8_cov_Ez}
\end{align}
where $f\sigma_{8,i} = f\sigma_8(z_i)$,
\begin{align}
I_i &= \int_0^{z_i} \frac{(1+z)^{3\gamma -1}}{E^{2\gamma}(z)} dz, \\
M_{ij} &= \int_0^{z_j} \frac{(1+z)^{3\gamma -1}}{H^{2\gamma}(z)} \frac{\text{Cov}[E(z), E_i]}{E(z) E_i} dz, \\
N_{ij} &= \int_0^{z_i} \int_0^{z_j} \frac{[(1+x)(1+y)]^{3\gamma -1}}{[E(x) E(y)]^{2\gamma}} \frac{\text{Cov}[E(x), E(y)]}{E(x) E(y)} dxdy.
\end{align}

In Fig.~\ref{fig:rec_fsig8}, we plot the variation of $f\sigma_8(z)$ reconstructed under different data combinations with respect to redshift $z$.
As shown in this figure, the uncertainties of the reconstructed $f\sigma_8(z)$ from BAO or CC+BAO data combination are much smaller than the observational uncertainties, which is due to the smaller variances of $H(z)$ from BAO data can significantly reduce the reconstructed errors of $f\sigma_8(z)$ at high redshifts.

One can find that the when the R18 prior are considered, the slopes of the reconstructed $f\sigma_8(z)$ varies greatly.
These changes are consistent with the reconstruction of $H(z)$ described in section~\ref{sec:rec_hz}.

\section{Results and discussion}
\label{sec:result}


\begin{table*}
\centering
\caption{68\% and 95\% confidence level constraints on the free parameters for various observational data combinations. Note that $H_0$ of the GP method is directly reconstructed from the OHD data. Here ``ALL'' correspond to the data combination of ``CC+BAO''. }
\label{tab:results}
\renewcommand\arraystretch{1.5}
\setlength{\tabcolsep}{3pt}
\begin{tabular}{cccccccc}
\hline\hline
Method & Data combination & $\sigma_8^0$ & $\Omega_m^0$ & $\gamma$ & $S_8$ & $H_0$ [km/s/Mpc] \\
\hline
\multirow{9}*{GP} &
RSD+CC+None
& $0.703^{+0.057}_{-0.16} < 0.969$ 
& $0.46^{+0.16}_{-0.10} {}^{+0.23}_{-0.27}$ 
& $0.91^{+0.19}_{-0.19} {}^{+0.38}_{-0.36}$
& $0.829^{+0.045}_{-0.039} {}^{+0.089}_{-0.086}$
& $67.36\pm4.8 \pm 9.6$ \\
~ & RSD+CC+R18
& $0.92^{+0.13}_{-0.24} {}^{+0.42}_{-0.33}$ 
& $0.169^{+0.057}_{-0.15} < 0.380$ 
& $0.48^{+0.10}_{-0.13} {}^{+0.23}_{-0.21}$
& $0.603^{+0.15}_{-0.052} {}^{+0.18}_{-0.27}$
& $73.24 \pm 1.6 \pm 3.2$ \\
~ & RSD+CC+P18 
& $0.754^{+0.068}_{-0.19} < 1.07$ 
& $0.41^{+0.19}_{-0.13} {}^{+0.27}_{-0.32}$
& $0.74^{+0.17}_{-0.17} {}^{+0.34}_{-0.32}$ 
& $0.815^{+0.062}_{-0.014} {}^{+0.10}_{-0.20}$
& $67.26\pm 0.60 \pm 1.20$ \\
\cline{2-7}
~ & RSD+BAO+None 
& $0.719^{+0.061}_{-0.16} {}^{+0.27}_{-0.22}$  
& $0.507^{+0.16}_{-0.095} {}^{+0.23}_{-0.28}$ 
& $1.01^{+0.20}_{-0.20} {}^{+0.38}_{-0.38}$ 
& $0.900^{+0.048}_{-0.041} {}^{+0.085}_{-0.094}$
& $63.22\pm 3.4 \pm 6.8$ \\
~ & RSD+BAO+R18
& $1.14^{+0.19}_{-0.19} > 0.828$ 
& $0.095^{+0.023}_{-0.077} {}^{+0.13}_{-0.089}$
& $0.463^{+0.079}_{-0.099} {}^{+0.18}_{-0.17}$ 
& $0.579^{+0.14}_{-0.082} {}^{+0.19}_{-0.23}$
& $73.15\pm 1.6 \pm 3.2$ \\
~ & RSD+BAO+P18 
& $0.92^{+0.12}_{-0.25} {}^{+0.43}_{-0.32} $
& $0.26^{+0.11}_{-0.18} {}^{+0.25}_{-0.23}$
& $0.62^{+0.13}_{-0.15} {}^{+0.27}_{-0.26}$ 
& $0.772^{+0.10}_{-0.026} {}^{+0.12}_{-0.21}$
& $67.22\pm 0.59 \pm 1.18$ \\
\cline{2-7}
~ & RSD+ALL+None
& $0.86^{+0.10}_{-0.23} {}^{+0.40}_{-0.31}$ 
& $0.32^{+0.14}_{-0.14} {}^{+0.25}_{-0.26} $
& $0.73^{+0.15}_{-0.18} {}^{+0.32}_{-0.31}$ 
& $0.817^{+0.068}_{-0.022} {}^{+0.11}_{-0.16}$
& $66.68\pm 3.1 \pm 6.2$ \\
~ & RSD+ALL+R18 
& $1.04^{+0.18}_{-0.26} {}^{+0.41}_{-0.34}$
& $0.123^{+0.035}_{-0.11} < 0.300$ 
& $0.460^{+0.083}_{-0.11} {}^{+0.21}_{-0.18}$ 
& $0.594^{+0.15}_{-0.067} {}^{+0.19}_{-0.24}$
& $72.62 \pm 1.5 \pm 3.0$ \\
~ & RSD+ALL+P18
& $0.91^{+0.12}_{-0.25} {}^{+0.42}_{-0.33}$
& $0.26^{+0.11}_{-0.18} {}^{+0.25}_{-0.25}$
& $0.61^{+0.13}_{-0.15} {}^{+0.30}_{-0.26}$ 
& $0.766^{+0.10}_{-0.022} {}^{+0.13}_{-0.22}$
& $67.25\pm 0.59 \pm 1.18$ \\
\hline
\multirow{9}*{$\Lambda$CDM} &
RSD+CC+None
& $0.80^{+0.13}_{-0.15} {}^{+0.25}_{-0.25}$
& $0.332^{+0.051}_{-0.070} {}^{+0.13}_{-0.11}$ 
& $0.68^{+0.28}_{-0.32} {}^{+0.56}_{-0.55}$ 
& $0.84^{+0.14}_{-0.20} {}^{+0.35}_{-0.31}$
& $67.7^{+3.1}_{-3.1} {}^{+5.9}_{-6.0}$ \\
~ & RSD+CC+R18
& $0.78^{+0.12}_{-0.15} {}^{+0.24}_{-0.24}$ 
& $0.261^{+0.032}_{-0.032} {}^{+0.066}_{-0.062}$ 
& $0.65^{+0.28}_{-0.33} {}^{+0.56}_{-0.55}$ 
& $0.73^{+0.12}_{-0.15} {}^{+0.26}_{-0.24}$
& $72.2\pm 1.4 \pm 2.8$ \\
~ & RSD+CC+P18 
& $0.80^{+0.12}_{-0.15} {}^{+0.25}_{-0.24}$
& $0.335^{+0.032}_{-0.032} {}^{+0.064}_{-0.062}$
& $0.67^{+0.27}_{-0.32} {}^{+0.56}_{-0.55}$
& $0.84^{+0.13}_{-0.16} {}^{+0.28}_{-0.26}$
& $67.31^{+0.59}_{-0.59} {}^{+1.1}_{-1.2}$ \\
\cline{2-7}
~ & RSD+BAO+None 
& $0.78^{+0.12}_{-0.15} {}^{+0.24}_{-0.24}$  
& $0.261^{+0.017}_{-0.020} {}^{+0.040}_{-0.035}$ 
& $0.65^{+0.28}_{-0.32} {}^{+0.55}_{-0.55}$
& $0.73\pm 0.11 {}^{+0.22}_{-0.21}$
& $70.2^{+1.2}_{-1.2} {}^{+2.3}_{-2.4}$ \\
~ & RSD+BAO+R18
& $0.77^{+0.12}_{-0.15} {}^{+0.24}_{-0.24}$ 
& $0.245^{+0.015}_{-0.015} {}^{+0.030}_{-0.028}$
& $0.64^{+0.27}_{-0.33} {}^{+0.56}_{-0.55}$
& $0.69^{+0.10}_{-0.13} {}^{+0.23}_{-0.21}$
& $71.31^{+0.93}_{-0.93} {}^{+1.8}_{-1.8}$ \\
~ & RSD+BAO+P18 
& $0.79^{+0.12}_{-0.15} {}^{+0.24}_{-0.24}$
& $0.297^{+0.012}_{-0.012} {}^{+0.025}_{-0.023}$
& $0.66^{+0.28}_{-0.28} {}^{+0.54}_{-0.54}$
& $0.81\pm 0.13 {}^{+0.25}_{-0.23}$
& $67.83^{+0.55}_{-0.55} {}^{+1.1}_{-1.1}$ \\
\cline{2-7}
~ & RSD+ALL+None
& $0.78^{+0.12}_{-0.15} {}^{+0.24}_{-0.24}$ 
& $0.267^{+0.017}_{-0.019} {}^{+0.035}_{-0.035}$
& $0.65^{+0.28}_{-0.32} {}5^{+0.55}_{-0.56}$
& $0.76\pm 0.13 {}^{+0.25}_{-0.24}$
& $69.9^{+1.1}_{-1.1} {}^{+2.1}_{-2.2}$ \\
~ & RSD+ALL+R18 
& $0.77^{+0.11}_{-0.15} {}^{+0.24}_{-0.23}$
& $0.251^{+0.014}_{-0.014} {}^{+0.029}_{-0.028}$ 
& $0.64^{+0.27}_{-0.33} {}^{+0.55}_{-0.54}$
& $0.72^{+0.11}_{-0.14} {}^{+0.25}_{-0.22}$
& $71.01^{+0.90}_{-0.90} {}^{+1.7}_{-1.8}$ \\
~ & RSD+ALL+P18
& $0.79^{+0.12}_{-0.15} {}^{+0.25}_{-0.24}$ 
& $0.299^{+0.012}_{-0.012} {}^{+0.023}_{-0.022}$
& $0.67^{+0.28}_{-0.33} {}^{+0.56}_{-0.55}$
& $0.79^{+0.12}_{-0.14} {}^{+0.26}_{-0.24}$
& $67.86\pm 0.53 \pm 1.0$ \\
\hline\hline
\end{tabular}
\end{table*}

\begin{figure}
\centering
\includegraphics[width=\linewidth]{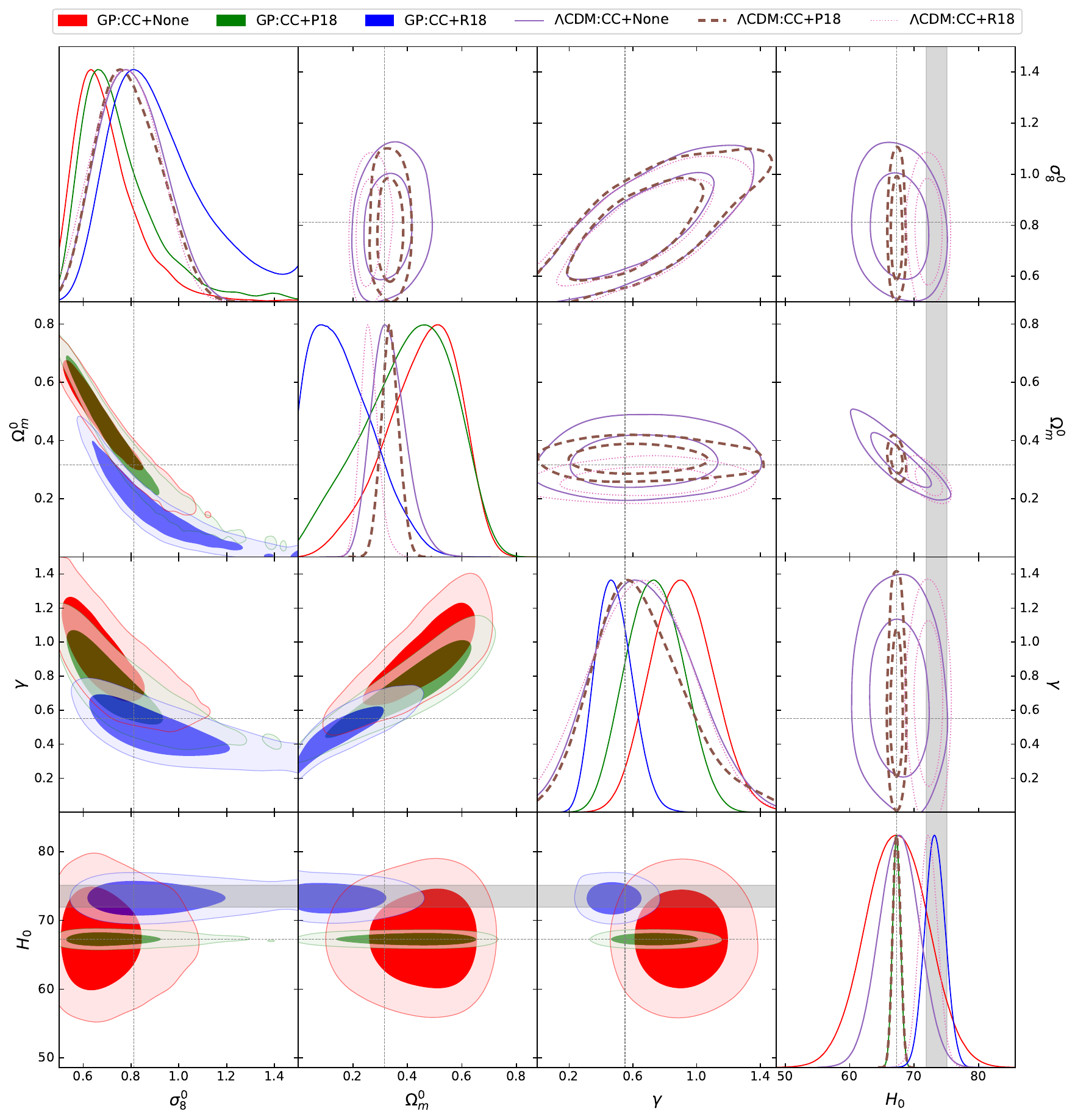}
\caption{The 68\% and 95\% confidence level contour plots 
of the free parameters 
and the corresponding one dimensional marginalized posterior distributions under the RSD+CC data combination. The vertical and horizontal dotted lines represent $\sigma_8^0 =0.8120$, $\Omega_m^0 = 0.3166$, $\gamma = 0.545$ and $H_0=67.27$. The grey vertical band corresponds to the R18 value.}
\label{fig:CC}
\end{figure}

\begin{figure}
\centering
\includegraphics[width=\linewidth]{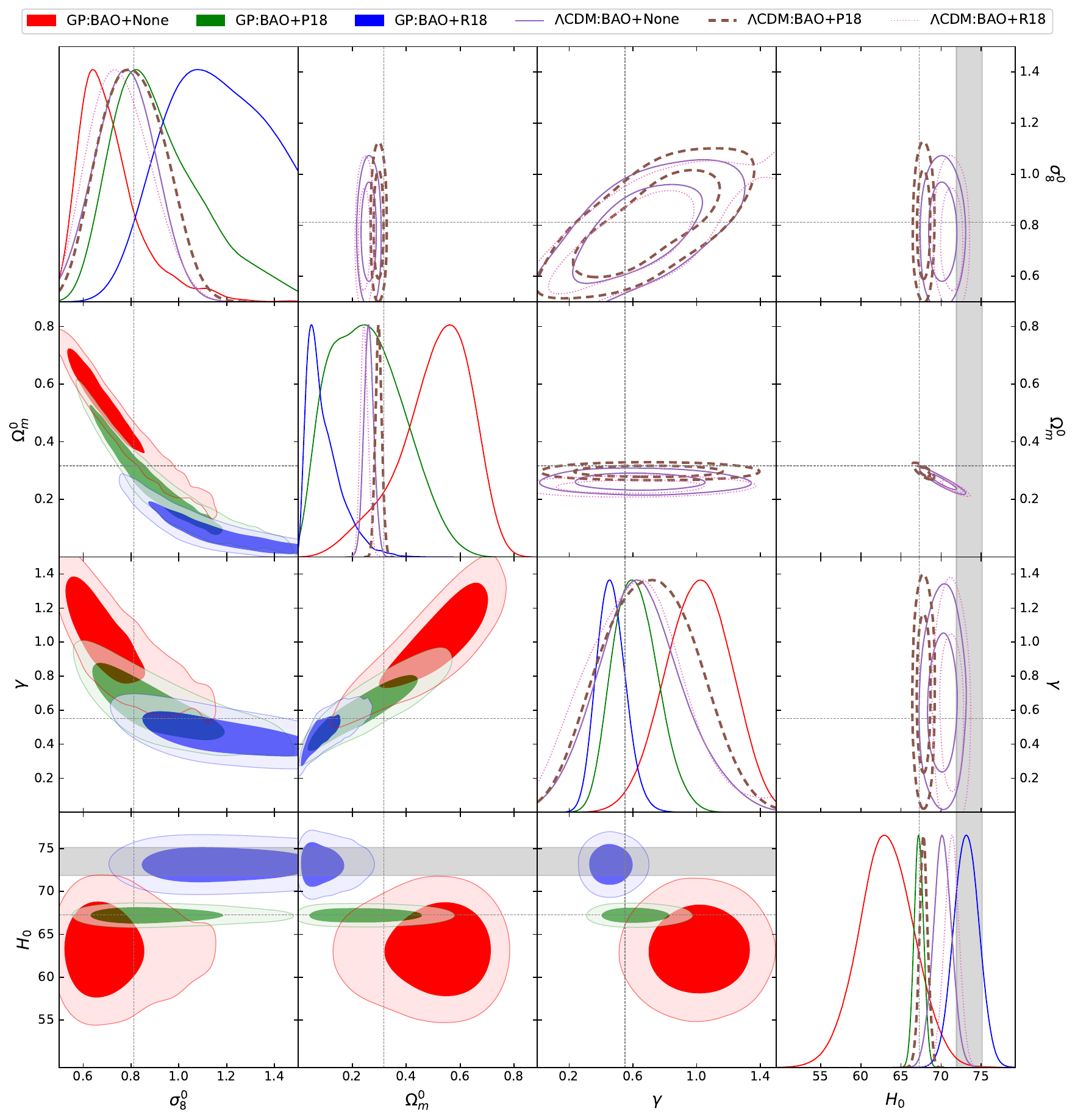}
\caption{The 68\% and 95\% confidence level contour plots 
of the free parameters 
and the corresponding one dimensional marginalized posterior distributions under the RSD+BAO data combination.. The vertical and horizontal dotted lines represent $\sigma_8^0 =0.8120$, $\Omega_m^0 = 0.3166$, $\gamma = 0.545$ and $H_0=67.27$. The grey vertical band corresponds to the R18 value.}
\label{fig:BAO}
\end{figure}

\begin{figure}
\centering
\includegraphics[width=\linewidth]{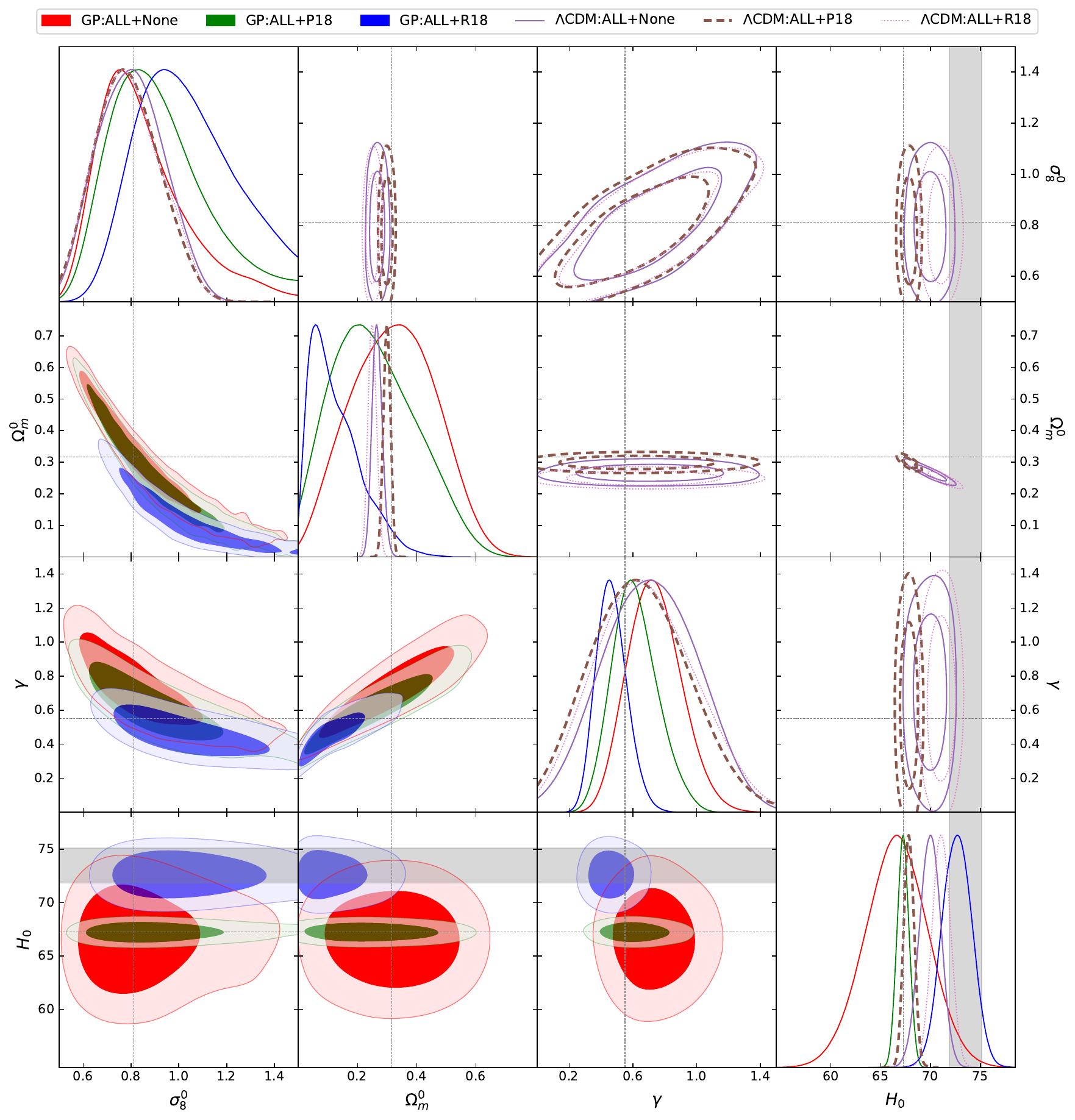}
\caption{The 68\% and 95\% confidence level contour plots 
of the free parameters 
and the corresponding one dimensional marginalized posterior distributions under the RSD+ALL data combination.. The vertical and horizontal dotted lines represent $\sigma_8^0 =0.8120$, $\Omega_m^0 = 0.3166$, $\gamma = 0.545$ and $H_0=67.27$. The grey vertical band corresponds to the R18 value.}
\label{fig:ALL}
\end{figure}

In this section, we describe the main results acquired from the model-independent method considered in this work.
As a comparison, we also give constrain on the $\Lambda$CDM model under the OHD and the RSD data points\footnote{For $\Lambda$CDM model, the parameterization of $f=\Omega_m(z)^{\gamma}$ are also used to constrain on cosmological parameters. And the cosntraint results of $\Lambda$CDM model are given under the combination of OHD data and $f\sigma_8(z)$ data. }.
The flat-linear priors for the free parameters we choose are $\sigma_8^0 \in [0.5, 1.5], \Omega_m^0 \in [0.0, 1.0]$, and $\gamma \in [0.0, 1.5]$.
For the $\Lambda$CDM model, we choose $H_0 \in [55, 90]$ and the priors of the other three parameters are the same as that of the GP method.

The summary of the observational constraints on the free parameters using different data combinations is listed in Table~\ref{tab:results}.
The results show that in the $\Lambda$CDM model, the different OHD data combinations and different priors of $H_0$ have little influence on $\sigma_8^0$ and $\gamma$, but have significant impacts on $\Omega_m^0$ and $H_0$.
Unlike that in the $\Lambda$CDM model, the various data combinations and the priors of $H_0$ have great influence on the constraints of all the free parameters of the GP method.

In Table~\ref{tab:results}, one can find that the $\Lambda$CDM model gives a much tighter constraint on $H_0$ than the GP method, this is because that, in the GP method, $H_0$ is reconstructed using only the OHD data.
By comparing our results of the $\Lambda$CDM model with the PLA results, where $\sigma_8^0 = 0.8120 \pm 0.0073$ and $\Omega_m^0 = 0.3166 \pm 0.0084$ \citep{Aghanim:2018eyx}, we find that $\sigma_8^0$ is consistent with the PLA results in $1\sigma$ region under different data combinations with or without prior on $H_0$.
Meanwhile, all the constraints of $\gamma$ in $\Lambda$CDM model are consistent with the GR prediction $\gamma = 0.55$ in $1\sigma$ region.

From Table~\ref{tab:results}, we note that in the $\Lambda$CDM model and the GP reconstruction method, the increasing of $H_0$ results in a decreasing of $\Omega_m^0$, which increases the tension level between our constraint $\Omega_m^0$ and that of the PLA results.
To better understand the relationship and tension level between the free parameters, in Figs.~\ref{fig:CC}, \ref{fig:BAO} and \ref{fig:ALL} we display the 2-D contour plots and the 1-D posterior distributions of the corresponding free parameters under different cases.
From the 1-D posterior distributions and the values in Table~\ref{tab:results}, one can find that in the GP method, the Hubble constant $H_0$ is smaller than that in the $\Lambda$CDM model when applying None or P19 prior on $H_0$, while $\Omega_m^0$ is larger and $\sigma_8^0$ is lower.
However, when a prior R18 is considered, we find the opposite trend.

The contour plots of the GP method in the three figures suggest that $\sigma_8^0$ is anti-correlated with $\Omega_m^0$ and $\gamma$, but $\Omega_m^0$ is positively correlated with $\gamma$, which are quite different from that of $\Lambda$CDM model.
As can be seen from Figs.~\ref{fig:CC} - \ref{fig:ALL} and the results in Table~\ref{tab:results}, the OHD data from BAO prefers larger $\sigma_8^0$, $\Omega_m^0$ and $\gamma$ than the CC data if we do not take any prior on $H_0$.
However, using all the OHD data points without prior on $H_0$, the best-fit of our results are closer to that of PLA results than using only OHD data from CC or BAO.
Comparing the results under the same data combination, it can be found that the three free parameters vary greatly under different priors of $H_0$, especially the matter density parameter $\Omega_m^0$, which suggests that $H_0$ has a great influence on constraining the three parameters.

\begin{figure}
\centering
\includegraphics[width=\linewidth]{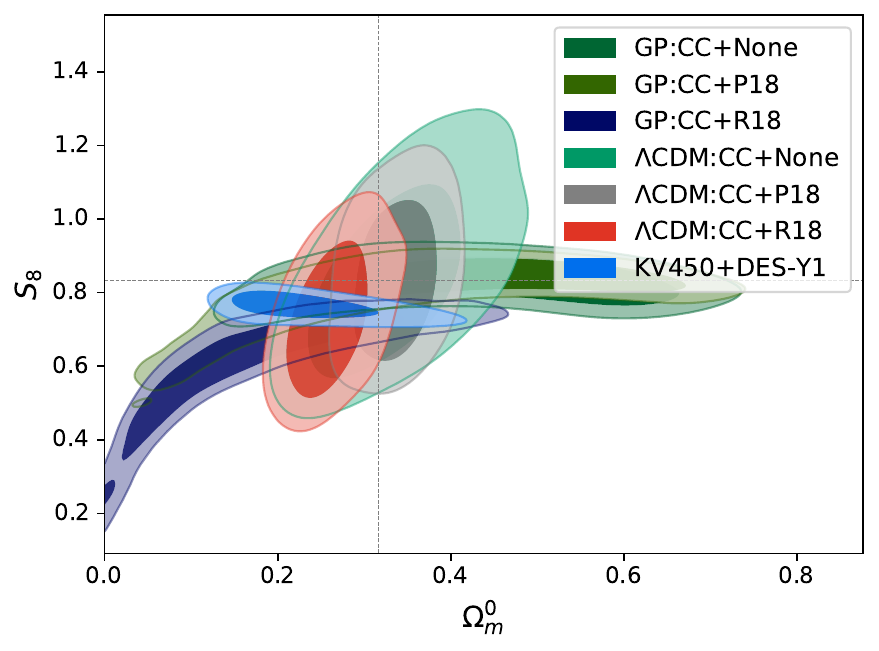}
\caption{68\% and 95\% confidence level contour plots for $S_8$ and $\Omega_m^0$ under RSD+CC data. The vertical and horizontal dotted lines represent $S_8 = 0.834$ and $\Omega_m^0 = 0.3166$ from PLA results.}
    \label{fig:s8cc}
\end{figure}

\begin{figure}
\centering
\includegraphics[width=\linewidth]{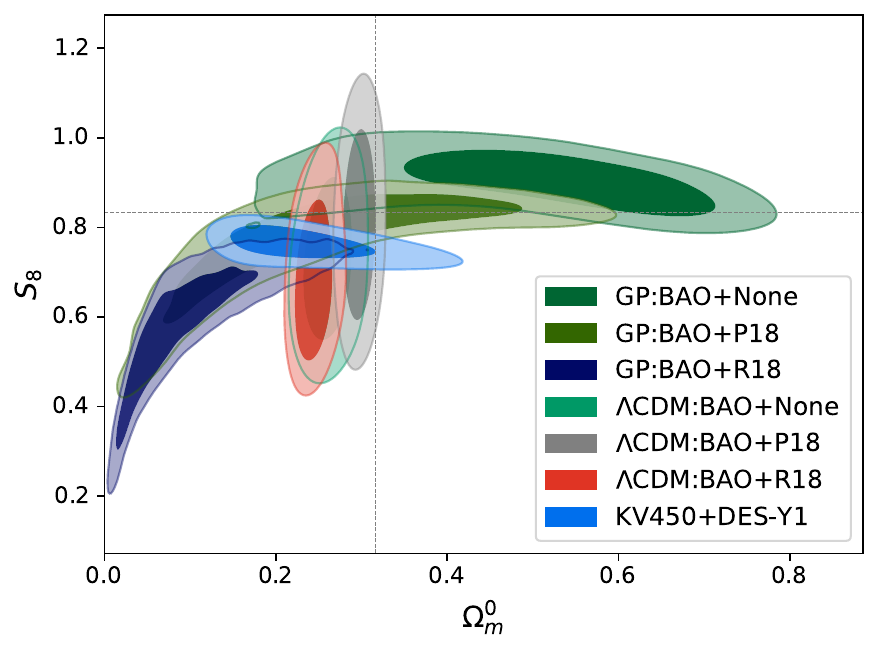}
\caption{68\% and 95\% confidence level contour plots for $S_8$ and $\Omega_m^0$ under RSD+BAO data. The vertical and horizontal dotted lines represent $S_8 = 0.834$ and $\Omega_m^0 = 0.3166$ from PLA results.}
    \label{fig:s8bao}
\end{figure}

\begin{figure}
\centering
\includegraphics[width=\linewidth]{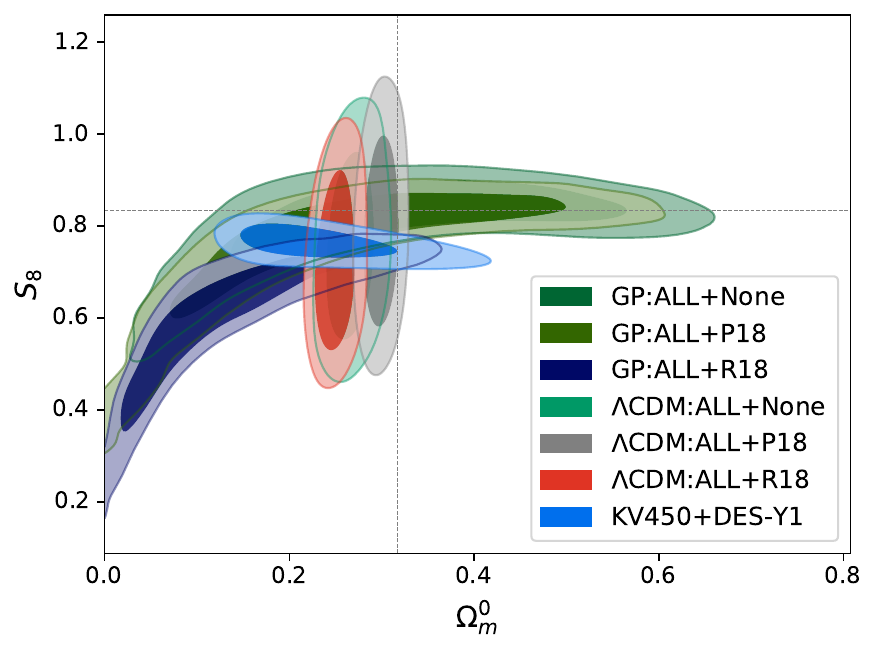}
\caption{68\% and 95\% confidence level contour plots for $S_8$ and $\Omega_m^0$ under RSD+ALL data. The vertical and horizontal dotted lines represent $S_8 = 0.834$ and $\Omega_m^0 = 0.3166$ from PLA results.}
    \label{fig:s8all}
\end{figure}

From the $\Omega_m^0 - \sigma_8^0$ contour plots in Figs.~\ref{fig:CC}, \ref{fig:BAO} and \ref{fig:ALL}, one can find that the tension level between our constraint $\Omega_m^0 -\sigma_8^0$ and the PLA values are within $1\sigma$ level when taking P18 $H_0$ prior and out of $2\sigma$ level when taking R18 prior.
The dashed lines in the $\Omega_m^0 - \sigma_8^0$ contour plots represent the best-fit values of PLA results.
On the other hand, the tension is often quantified using the $S_8 \equiv \sigma_8^0 \sqrt{\Omega_m^0 /0.3}$ parameter, along the main degeneracy direction of weak lensing measurements \citep{DiValentino:2020vvd}.
Thus, the contour plots of $S_8 - \Omega_m^0$ of different cases are shown in Fig.~\ref{fig:s8cc} - \ref{fig:s8all}.
In the $S_8 - \Omega_m^0$ planes, the result from KV450 and DES-Y1 measurements \citep{Joudaki:2019pmv} are plotted as a contrast.
From these contourplots, one can find that the correlations between $S_8$ and $\Omega_m^0$ are quite different between GP method and $\Lambda$CDM model.
One can also find that $S_8$ is positively correlated with $\Omega_m^0$ in our methods, which is different with the KV450+DES-Y1 results.
This suggest that our methods are helpful in breaking the degeneracies of the two parameters if the KV450 and DES-Y1 data are considered.
Meanwhile, from the contours, one can obtain that the difference of $S_8$ in the GP method is less than $2\sigma$ with PLA results when applying None and P19 priors of $H_0$.
But when the R19 prior are applied, the tension is bigger than $2\sigma$.

Besides, though our constrainton the growth index $\gamma$ varies greatly in different cases, the tension level between our constraint result and the GR prediction are almost less than $2\sigma$.

\section{Summary}
\label{sec:sum}

In this work, we consider the constraints on the matter fluctuation amplitude $\sigma_8^0$, the matter density parameter $\Omega_m^0$, and the growth index $\gamma$ by using the latest OHD and RSD data combinations.
To be model-independent, we use GP method to reconstruct the Hubble parameter $H(z)$ from different OHD data combinations, and then obtain a theoretical $f\sigma_8(z)$ function. 
We then use the reconstructed $f\sigma_8(z)$ and the RSD data to sample the free parameters by means of the MCMC method.
Here, to reduce the impact of the AP effect, we also correct the RSD data using the $H(z)$ and $D_A(z)$ reconstructed by GP method.

From the curves of the reconstructed $H(z)$ and $E(z)$ shown in Figs.~\ref{fig:rec_hz} and \ref{fig:rec_Ez} respectively, one can find that the expansion history varies greatly from the $\Lambda$CDM model when taking R18 prior on $H_0$, especially at the low redshifts.
This indicates that to solve or relax the $H_0$ tension problem, it is necessary to have a model with a different expansion history from the $\Lambda$CDM model or a dynamical dark energy model which is different from the cosmological constant at low redshifts.

As in Fig.~\ref{fig:rec_fsig8}, the reconstructed $f\sigma_8(z)$ fits well with the $\Lambda$CDM model when using the same cosmological parameters. 
Meanwhile, we also find that both the expansion history $H(z)$ and the prior of $H_0$ have great influences on the reconstruction of the $f\sigma_8(z)$, which is also supported by the MCMC sampling results obtained in this paper.

Our results show that the tension level between our best-fit of $\Omega_m^0 - \sigma_8^0$ and that of the PLA results are no more than $2\sigma$ in the cases of no $H_0$ prior.
And the tension will even disappear (because of less than $1\sigma$) when adopting P18 $H_0$ prior.
This conclusion are also supported by the $S_8 -\Omega_m^0$ contour plots.
Meanwhile, the different correlations between our results and the KV450+DES-Y1 results suggest that using our method and combine the RSD and OHD data may be helpful to break the $S_8 - \Omega_m^0$ degeneracies in cosmic shear measurements.
However, one should note that our constraints on $\Omega_m^0$ and $\sigma_8^0$ are much looser compared with the PLA results, which means that the small tension level may be caused by the larger uncertainties. 
Nonetheless, using all the OHD data points and the RSD data without $H_0$ prior, our constraint result of $\Omega_m^0$ and $\sigma_8^0$ are very close to the PLA values, and the tension level between the growth index $\gamma$ and the GR prediction $\gamma = 0.55$ is at $1\sigma$ level.

From our analysis, one can find that the mean values of $H_0$ have a great influence on the reconstructed expanding history $H(z)$, the reconstructed growth history $f\sigma_8(z)$, and the constraint results of cosmological parameters.
Compared with R18, \cite{Riess:2019cxk} present an even larger value of $H_0$, i.e., $74.03 \pm 1.42$ km/s/Mpc (R19).
Thus, if this value is chosen as $H_0$ prior, the reconstructed $H(z)$ will be much bigger at low redshifts. 
According to the results shown in Table~\ref{tab:results}, $H_0$ is positively-correlated with $\sigma_8^0$ and anti-correlated with $\Omega_m^0$ and $\gamma$, thus, if the R19 is used, we will get a larger $\sigma_8^0$ but smaller $\Omega_m^0$ and $\gamma$.

It should be noted that the possibility of correlations exists in the $f\sigma_8$ datapoints are not considered in the present paper.
\cite{Kazantzidis:2018rnb} had considered positive correlations in 12 randomly selected pairs of the 63 $f\sigma_8(z)$ datapoints.
They find that the introduction of a nontrivial covariance matrix does not change the qualitative conclusions of their analysis of the tensions.
So in this paper, we only take the correlation matrix generated from the fiducial model corrections (see section \ref{sec:rec_fs8_ap}).

\section*{Acknowledgments}

This work is supported in part by National Natural Science Foundation of China under Grant No. 11675032 and Grant No. 12075042 (People's Republic of China).
E.-K. Li thanks J. Geng for suggestions of English language revising.

\section*{Data availability}

The data underlying this article are available in the article and the references details can be found in the APPENDIX A.




\bibliographystyle{mnras}


\appendix


\section{$f\sigma_8$ growth data and observational Hubble data combinations}

\begin{table*}
    \caption{A compilation of RSD data that reported in different surveys.}
    \label{tab:fsigma8_data}
    \begin{centering}
        \scriptsize 
        \renewcommand\arraystretch{1}
        \begin{tabular}{|c|c|c|c|c|c|c|}
            \hline
            Index & Dataset & $z$ & $f\sigma_8(z)$ & Refs. & Year & Fiducial Cosmology \\
            \hline
            1 & SDSS-LRG & $0.35$ & $0.440\pm 0.050$ & [1] &  30 October 2006 &$(\Omega_{0m},\Omega_K,\sigma_8$ [2])$=(0.25,0,0.756)$ \\
            2 & VVDS & $0.77$ & $0.490\pm 0.18$ & [1]  & 6 October 2009 & $(\Omega_{0m},\Omega_K,\sigma_8)=(0.25,0,0.78)$ \\
            3 & 2dFGRS & $0.17$ & $0.510\pm 0.060$ & [1]  &  6 October 2009 & $(\Omega_{0m},\Omega_K)=(0.3,0,0.9)$ \\
            4 & 2MRS &0.02& $0.314 \pm 0.048$ &  [3], [4] & 13 Novemver 2010 & $(\Omega_{0m},\Omega_K,\sigma_8)=(0.266,0,0.65)$ \\
            5 & SnIa+IRAS &0.02& $0.398 \pm 0.065$ & [5], [4] & 20 October 2011 & $(\Omega_{0m},\Omega_K,\sigma_8)=(0.3,0,0.814)$\\
            6 & SDSS-LRG-200 & $0.25$ & $0.3512\pm 0.0583$ & [6] & 9 December 2011 & $(\Omega_{0m},\Omega_K,\sigma_8)=(0.276,0,0.8)$  \\
            7 & SDSS-LRG-200 & $0.37$ & $0.4602\pm 0.0378$ & [6] & 9 December 2011 & \\
            8 & SDSS-LRG-60 & $0.25$ & $0.3665\pm0.0601$ & [6] & 9 December 2011 & $(\Omega_{0m},\Omega_K,\sigma_8)=(0.276,0,0.8)$ \\
            9 & SDSS-LRG-60 & $0.37$ & $0.4031\pm0.0586$ & [6] & 9 December 2011 &\\
            10 & WiggleZ & $0.44$ & $0.413\pm 0.080$ & [7] & 12 June 2012  & $(\Omega_{0m},h,\sigma_8)=(0.27,0.71,0.8)$ \\
            11 & WiggleZ & $0.60$ & $0.390\pm 0.063$ & [7] & 12 June 2012 &  $C_{ij}=Eq.~\eqref{eq:Cij_WiggleZ}$\\
            12 & WiggleZ & $0.73$ & $0.437\pm 0.072$ & [7] & 12 June 2012 &\\
            13 & 6dFGS& $0.067$ & $0.423\pm 0.055$ & [8] & 4 July 2012 & $(\Omega_{0m},\Omega_K,\sigma_8)=(0.27,0,0.76)$ \\
            14 & SDSS-BOSS& $0.30$ & $0.407\pm 0.055$ & [9] & 11 August 2012 & $(\Omega_{0m},\Omega_K,\sigma_8)=(0.25,0,0.804)$ \\
            15 & SDSS-BOSS& $0.40$ & $0.419\pm 0.041$ & [9] & 11 August 2012 & \\
            16 & SDSS-BOSS& $0.50$ & $0.427\pm 0.043$ & [9] & 11 August 2012 & \\
            17 & SDSS-BOSS& $0.60$ & $0.433\pm 0.067$ & [9] & 11 August 2012 & \\
            18 & Vipers& $0.80$ & $0.470\pm 0.080$ & [10] & 9 July 2013 & $(\Omega_{0m},\Omega_K,\sigma_8)=(0.25,0,0.82)$  \\
            19 & SDSS-DR7-LRG & $0.35$ & $0.429\pm 0.089$ & [11]  & 8 August 2013 & $(\Omega_{0m},\Omega_K,\sigma_8$ [12])$=(0.25,0,0.809)$\\
            20 & GAMA & $0.18$ & $0.360\pm 0.090$ & [13] & 22 September 2013 & $(\Omega_{0m},\Omega_K,\sigma_8)=(0.27,0,0.8)$ \\
            21& GAMA & $0.38$ & $0.440\pm 0.060$ & [13]  & 22 September 2013 & \\
            22 & BOSS-LOWZ& $0.32$ & $0.384\pm 0.095$ & [14]  & 17 December 2013  & $(\Omega_{0m},\Omega_K,\sigma_8)=(0.274,0,0.8)$ \\
            23 & SDSS DR10 and DR11 & $0.32$ & $0.48 \pm 0.10$ & [14] &   17 December 2013 & $(\Omega_{0m},\Omega_K,\sigma_8$ [15])$=(0.274,0,0.8)$ \\
            24 & SDSS DR10 and DR11 & $0.57$ & $0.417 \pm 0.045$ & [14] &  17 December 2013 &  \\
            25 & SDSS-MGS & $0.15$ & $0.490\pm0.145$ & [16] & 30 January 2015 & $(\Omega_{0m},h,\sigma_8)=(0.31,0.67,0.83)$ \\
            26 & SDSS-veloc & $0.10$ & $0.370\pm 0.130$ & [17]  & 16 June 2015 & $(\Omega_{0m},\Omega_K,\sigma_8$ [18])$=(0.3,0,0.89)$ \\
            27 & FastSound& $1.40$ & $0.482\pm 0.116$ & [19]  & 25 November 2015 & $(\Omega_{0m},\Omega_K,\sigma_8$ [20])$=(0.27,0,0.82)$\\
            28 & SDSS-CMASS & $0.59$ & $0.488\pm 0.060$ & [21] & 8 July 2016 & $\ \ (\Omega_{0m},h,\sigma_8)=(0.307115,0.6777,0.8288)$ \\
            29 & BOSS DR12 & $0.38$ & $0.497\pm 0.045$ & [22] & 11 July 2016 & $(\Omega_{0m},\Omega_K,\sigma_8)=(0.31,0,0.8)$ \\
            30 & BOSS DR12 & $0.51$ & $0.458\pm 0.038$ & [22] & 11 July 2016 & \\
            31 & BOSS DR12 & $0.61$ & $0.436\pm 0.034$ & [22] & 11 July 2016 & \\
            32 & BOSS DR12 & $0.38$ & $0.477 \pm 0.051$ & [23] & 11 July 2016 & $(\Omega_{0m},h,\sigma_8)=(0.31,0.676,0.8)$ \\
            33 & BOSS DR12 & $0.51$ & $0.453 \pm 0.050$ & [23] & 11 July 2016 & \\
            34 & BOSS DR12 & $0.61$ & $0.410 \pm 0.044$ & [23] & 11 July 2016 &  \\
            35 & Vipers v7& $0.76$ & $0.440\pm 0.040$ & [24] & 26 October 2016  & $(\Omega_{0m},\sigma_8)=(0.308,0.8149)$ \\
            36 & Vipers v7 & $1.05$ & $0.280\pm 0.080$ & [24] & 26 October 2016 &\\
            37 & BOSS LOWZ & $0.32$ & $0.427\pm 0.056$ & [25] & 26 October 2016 & $(\Omega_{0m},\Omega_K,\sigma_8)=(0.31,0,0.8475)$\\
            38 & BOSS CMASS & $0.57$ & $0.426\pm 0.029$ & [25] & 26 October 2016 & \\
            39 & Vipers  & $0.727$ & $0.296 \pm 0.0765$ & [26] &  21 November 2016 & $(\Omega_{0m},\Omega_K,\sigma_8)=(0.31,0,0.7)$\\
            40 & 6dFGS+SnIa & $0.02$ & $0.428 \pm 0.0465$ & [27] & 29 November 2016 & $(\Omega_{0m},h,\sigma_8)=(0.3,0.683,0.8)$ \\
            41 & Vipers  & $0.6$ & $0.48 \pm 0.12$ & [28] & 16 December 2016 & $(\Omega_{0m},\Omega_b,n_s,\sigma_8$ [29])= $(0.3, 0.045, 0.96,0.831)$\\
            42 & Vipers  & $0.86$ & $0.48 \pm 0.10$ & [28] & 16 December 2016  & \\
            43 & Vipers PDR-2& $0.60$ & $0.550 \pm 0.120$ & [30] & 16 December 2016 & $(\Omega_{0m},\Omega_b,\sigma_8)=(0.3,0.045,0.823)$ \\
            44 & Vipers PDR-2& $0.86$ & $0.400 \pm 0.110$ & [30] & 16 December 2016 &\\
            45 & SDSS DR13  & $0.1$ & $0.48 \pm 0.16$ & [31] & 22 December 2016 & $(\Omega_{0m},\sigma_8$ [18])$=(0.25,0.89)$ \\
            46 & 2MTF & 0.001 & $0.505 \pm 0.085$ &  [32] & 16 June 2017 & $(\Omega_{0m},\sigma_8)=(0.3121,0.815)$\\
            47 & Vipers PDR-2 & $0.85$ & $0.45 \pm 0.11$ & [33] & 31 July 2017  &  $(\Omega_b,\Omega_{0m},h)=(0.045,0.30,0.8)$ \\
            48 & BOSS DR12 & $0.31$ & $0.469 \pm 0.098$ & [34]  & 15 September 2017 & $(\Omega_{0m},h,\sigma_8)=(0.307,0.6777,0.8288)$\\
            49 & BOSS DR12 & $0.36$ & $0.474 \pm 0.097$ & [34]  & 15 September 2017 & \\
            50 & BOSS DR12 & $0.40$ & $0.473 \pm 0.086$ & [34]  & 15 September 2017 & \\
            51 & BOSS DR12 & $0.44$ & $0.481 \pm 0.076$ & [34]  & 15 September 2017 & \\
            52 & BOSS DR12 & $0.48$ & $0.482 \pm 0.067$ & [34]  & 15 September 2017 & \\
            53 & BOSS DR12 & $0.52$ & $0.488 \pm 0.065$ & [34]  & 15 September 2017 & \\
            54 & BOSS DR12 & $0.56$ & $0.482 \pm 0.067$ & [34]  & 15 September 2017 & \\
            55 & BOSS DR12 & $0.59$ & $0.481 \pm 0.066$ & [34]  & 15 September 2017 & \\
            56 & BOSS DR12 & $0.64$ & $0.486 \pm 0.070$ & [34]  & 15 September 2017 & \\
            57 & SDSS DR7 & $0.1$ & $0.376 \pm 0.038$ & [35] & 12 December 2017 & $(\Omega_{0m},\Omega_b,\sigma_8)=(0.282,0.046,0.817)$ \\
            58 & SDSS-IV & $1.52$ & $0.420 \pm 0.076$ &  [36] & 8 January 2018  & $(\Omega_{0m},\Omega_b h^2,\sigma_8)=(0.26479, 0.02258,0.8)$ \\ 
            59 & SDSS-IV & $1.52$ & $0.396 \pm 0.079$ & [37] & 8 January 2018 & $(\Omega_{0m},\Omega_b h^2,\sigma_8)=(0.31,0.022,0.8225)$ \\ 
            60 & SDSS-IV & $0.978$ & $0.379 \pm 0.176$ & [38]  & 9 January 2018 &$(\Omega_{0m},\sigma_8)=(0.31,0.8)$\\
            61 & SDSS-IV & $1.23$  & $0.385 \pm 0.099$ & [38]  & 9 January 2018 & \\
            62 & SDSS-IV & $1.526$ & $0.342 \pm 0.070$ & [38]  & 9 January 2018 & \\
            63 & SDSS-IV & $1.944$ & $0.364 \pm 0.106$ & [38]  & 9 January 2018 & \\
            \hline
        \end{tabular}\par
        Reference: 
        [1] \cite{Song:2008qt},
        [2] \cite{Tegmark:2006az}, 
        [3] \cite{Davis:2010sw}, 
        [4] \cite{Hudson:2012gt}, 
        [5] \cite{Turnbull:2011ty}, 
        [6] \cite{Samushia:2011cs}, 
        [7] \cite{Blake:2012pj}, 
        [8] \cite{Beutler:2012px}, 
        [9] \cite{Tojeiro:2012rp}, 
        [10] \cite{delaTorre:2013rpa}, 
        [11] \cite{Chuang:2012qt}, 
        [12] \cite{Komatsu:2010fb},
        [13] \cite{Blake:2013nif},
        [14] \cite{Sanchez:2013tga},
        [15] \cite{Anderson:2013zyy},
        [16] \cite{Howlett:2014opa},
        [17] \cite{Feix:2015dla},
        [18] \cite{Tegmark:2003uf}, 
        [19] \cite{Okumura:2015lvp}, 
        [20] \cite{Hinshaw:2012aka},
        [21] \cite{Chuang:2013wga},
        [22] \cite{Alam:2016hwk},
        [23] \cite{Beutler:2016arn},
        [24] \cite{Wilson:2016ggz},
        [25] \cite{Gil-Marin:2016wya},
        [26] \cite{Hawken:2016qcy}, 
        [27] \cite{Huterer:2016uyq}, 
        [28] \cite{delaTorre:2016rxm}, 
        [29] \cite{Ade:2015xua}, 
        [30] \cite{Pezzotta:2016gbo}, 
        [31] \cite{Feix:2016qhh}, 
        [32] \cite{Howlett:2017asq}, 
        [33] \cite{Mohammad:2017lzz}, 
        [34] \cite{Wang:2017wia}, 
        [35] \cite{Shi:2017qpr}, 
        [36] \cite{Gil-Marin:2018cgo}, 
        [37] \cite{Hou:2018yny}, 
        [38] \cite{Zhao:2018jxv}.
    \end{centering}
\end{table*}


\begin{table}
\centering
\caption{The latest Hubble parameter measurements $H(z)$ (in units of km\,s${}^{-1}$\,Mpc${}^{-1}$) and their errors $\sigma_H$ at redshift $z$ obtained from the differential age method (DA).}
\label{tab:Hubble_CC}
\begin{tabular}{|c|c|c|c|c|}
\hline
Index  & $z$ & $H(z)$ & $\sigma_H$ & Reference \\
\hline
1  & 0.07   &  69.0  &  19.6 & \multirow{4}*{\cite{Zhang:2012mp}} \\
2  & 0.12   &  68.6  &  26.2 & \\
3  & 0.2&  72.9  &  29.6 & \\
4  & 0.28   &  88.8  &  36.6 & \\
\hline
5  & 0.1&  69&  12   & \multirow{11}*{\cite{Stern:2009ep}} \\
6  & 0.17   &  83&  8& \\
7  & 0.27   &  77&  14   & \\
8  & 0.4&  95&  17   & \\
9  & 0.48   &  97&  60   & \\
10 & 0.88   &  90&  40   & \\
11 & 0.9&  117   &  23   & \\
12 & 1.3&  168   &  17   & \\
13 & 1.43   &  177   &  18   & \\
14 & 1.53   &  140   &  14   & \\
15 & 1.75   &  202   &  40   & \\
\hline
16 & 0.1797 & 75 & 4 & \multirow{8}*{\cite{Moresco:2012jh}} \\
17 & 0.1993 & 75 & 5 & \\
18 & 0.3519 & 83 & 14& \\
19 & 0.5929 & 104& 13& \\
20 & 0.6797 & 92 & 8 & \\
21 & 0.7812 & 105& 12& \\
22 & 0.8754 & 125& 17& \\
23 & 1.037  & 154& 20& \\
\hline
24 & 0.3802 & 83 & 13.5  & \multirow{5}*{\cite{Moresco:2016mzx}} \\
25 & 0.4004 & 77 & 10.2  & \\
26 & 0.4247 & 87.1   & 11.2  & \\
27 & 0.4497 & 92.8   & 12.9  & \\
28 & 0.4783 & 80.9   & 9 & \\
\hline
29 & 1.363  & 160& 33.6  & \multirow{2}*{\cite{Moresco:2015cya}} \\
30 & 1.965  & 186.5  & 50.4  & \\
\hline
31 & 0.47   & 89 & 34& \cite{Ratsimbazafy:2017vga} \\
\hline
\end{tabular}
\end{table}

\begin{table}
\centering
\caption{The latest Hubble parameter measurements $H(z)$ (in units of km\,s${}^{-1}$\,Mpc${}^{-1}$) and their errors $\sigma_H$ at redshift $z$ obtained from the radial BAO method (clustering).}
\label{tab:Hubble_BAO}
\begin{tabular}{|c|c|c|c|c|}
\hline
Index & $z$ & $H(z)$ & $\sigma_H$ & Reference \\
\hline
1  & 0.24   & 79.69  & 2.65 & \multirow{2}*{\cite{Gaztanaga:2008xz}} \\
2  & 0.43   & 86.45  & 3.68 & \\
\hline
3  & 0.3& 81.7   & 6.22 & \cite{Oka:2013cba} \\
\hline
4  & 0.31   & 78.17  & 4.74 & \multirow{6}*{\cite{Wang:2016wjr}} \\
5  & 0.36   & 79.93  & 3.39 & \\
6  & 0.40   & 82.04  & 2.03 & \\
7  & 0.44   & 84.81  & 1.83 & \\
8  & 0.48   & 87.79  & 2.03 & \\
9  & 0.52   & 94.35  & 2.65 & \\
10 & 0.56   & 93.33  & 2.32 & \\
11 & 0.59   & 98.48  & 3.19 & \\
12 & 0.64   & 98.82  & 2.99 & \\
\hline
13 & 0.35   & 82.7   & 8.4  & \cite{Chuang:2012qt} \\
\hline
14 & 0.38   & 81.5   & 1.9  & \multirow{3}*{\cite{Alam:2016hwk}} \\
15 & 0.51   & 90.4   & 1.9  & \\
16 & 0.61   & 97.3   & 2.1  & \\
\hline
17 & 0.44   & 82.6   & 7.8  & \multirow{3}*{\cite{Blake:2012pj}} \\
18 & 0.6& 87.9   & 6.1  & \\
19 & 0.73   & 97.3   & 7& \\
\hline
20 & 0.57   & 96.8   & 3.4  & \cite{Anderson:2013zyy} \\
\hline
21 & 2.33   & 224& 8& \cite{Bautista:2017zgn} \\
\hline
22 & 2.34   & 222& 7& \cite{Delubac:2014aqe} \\
\hline
23 & 2.36   & 226& 8& \cite{Font-Ribera:2013wce} \\
\hline
\end{tabular}
\end{table}

\bsp	
\label{lastpage}
\end{document}